\documentclass[a4paper,12pt]{article}
\usepackage[dvips]{color}
\usepackage{ulem}
\usepackage{multirow}
\usepackage{latexsym}
\usepackage{epsfig,epsf,subfig,graphicx,amsfonts,amssymb,color, subfig, color}
\usepackage[english]{babel}
\usepackage[utf8]{inputenc}
\usepackage{amsmath,bm,amsfonts,amssymb,upgreek}     
\usepackage{graphicx}                       
\usepackage{xfrac}                          
\usepackage{grffile}                        
\usepackage{xcolor}
\usepackage{color}
\usepackage{upgreek}
\usepackage{physics}
\usepackage[utf8]{inputenc}
\usepackage[colorlinks, citecolor=blue, urlcolor=blue, linkcolor=blue]{hyperref}

\usepackage{amsthm}
\usepackage{graphicx}
\usepackage{dcolumn}
\usepackage{ulem}
\usepackage[utf8]{inputenc}
\usepackage[T1]{fontenc}
\usepackage{mathptmx}
\usepackage{etoolbox}
\usepackage{xcolor}
\usepackage{multirow}
\usepackage{upgreek}%
\usepackage{mathrsfs}%
\usepackage{color}%
\usepackage{textcomp}%
\usepackage{manyfoot}%
\usepackage{booktabs}%
\usepackage{algorithm}%
\usepackage{algorithmicx}%
\usepackage{algpseudocode}%
\usepackage{listings}%

\usepackage[T1]{fontenc}

\title{Quantifying multifractal anisotropy \\in two dimensional objects}
\author{Rafał Rak$^1$, Stanisław Drożdż$^{2,3}$, Jarosław Kwapień$^2$, Paweł Oświęcimka$^2$}

\begin{document}
\maketitle
\noindent
\begin{center}
$^1$\textit{College of Natural Sciences, University of Rzeszów, Pigonia 1, 35-310 Rzeszów, Poland}\\
$^2$\textit{Complex Systems Theory Department, Institute of Nuclear Physics, Polish Academy of Sciences, ul.~Radzikowskiego 152, 31-342 Krak\'ow, Poland}\\
$^3$\textit{Faculty of Computer Science and Telecommunications, Cracow University of Technology, ul.~Warszawska 24, 31-155 Krak\'ow, Poland}\\

\end{center}

\begin{abstract}
An efficient method of exploring the effects of anisotropy in the fractal properties of 2D surfaces and images is proposed. It can be viewed as a direction-sensitive generalization of the multifractal detrended fluctuation analysis (MFDFA) into 2D. It is tested on synthetic structures to ensure its effectiveness, with results indicating consistency. The interdisciplinary potential of this method in describing real surfaces and images is demonstrated, revealing previously unknown directional multifractality in data sets from the Martian surface and the Crab Nebula. The multifractal characteristics of Jackson Pollock's paintings are also analyzed. The results point to their evolution over the time of creation of these works.
\end{abstract}

\textbf{Keywords}: Complex systems, Fractal analysis, 2D MFDFA, Rough surfaces,\\ Anisotropy.

\maketitle

\begin{quotation}
Multifractality is a pivotal concept within the science of complexity as it transcends traditional disciplinary boundaries and intersects with different disciplines. It thus provides a unified framework for quantifying complexity across a diverse range of fields. Its ability to capture the intricate and heterogeneous scaling behaviors found in many systems makes it an extremely valuable tool in various scientific and applied domains.
At the same time, multifractality is a mathematically advanced concept, and its conversion into practical and reliable algorithms is a delicate matter. So far, satisfactory algorithms have been developed and validated for one-dimensional problems, most often of the time series type. In general, two-dimensional problems common in nature pose disproportionately greater numerical challenges. A practical algorithm is proposed here, which is constructed in such a way that it captures those effects that may only appear in more than one-dimensional problems. These include, in particular, the effects of spatial anisotropy, asymmetry or directionality.

\end{quotation}

\section{Introduction}
\label{sect::introduction}

Like many objects in Nature, rough surfaces or patchy images often reveal scaling whose quantitative characteristics vary with their point-to-point fluctuation magnitude. This feature can be described in terms of multifractality, i.e., the property that requires multiple scaling exponents and multiple fractal dimensions to describe a given object~\cite{HalseyTC-1986a,MeakinP-1987a,AtamspacherA-1989a}. From a practical perspective, it is especially important to have reliable tools of quantifying such effects for data sets containing values of the measured observables that are uniformly sampled in time or in space. In the former case, several approaches have been developed over the years to extract fractal properties of 1-dimensional time series~\cite{HurstHE-1951a,ArneodoA-1995a,PengCK-1994a,KantelhardtJ-2002a,AlessioE-2002a,DiMatteoT-2007,GuGF-2010a,JiangZQ-2019}, the most widely applied of which is multifractal detrended fluctuation analysis (MFDFA)~\cite{KantelhardtJ-2002a, Gebarowski-2019ND, rak2, rak3, Oswiecimka-2005PA, Rak-2018PA, Watorek-2021PR, Dutta-2013FP, drozdz-2016IS, Watorek-2024BSPC, Sierra-2024CSF, Oswiecimka-2024ND, Hajian-2010PA,Ausloos-2012PRE, Wang-2023F, Klamut-2020PRE, Gulich-2012G, Chanda-2020FNL, Das-2014SR} due to its relatively good reliability~\cite{OswiecimkaP-2006a,ShaoYH-2012}. In the latter case, which requires considering at least 2-dimensional data arrays, the spectrum of available tools is poorer~\cite{MandelbrotBB-1967a,GrassbergerP-1983a,ArneodoA-2000a}. However, as the need for analysis of rough surfaces~\cite{ThomasTR-1999a,WuJJ-2000a,WaechterM-2004a,PerssonBNJ-2014a} and 2-dimensional images~\cite{BlacherS-1993a,EmersonCW-1999a,KuikkaJT-2002a,LiuY-2020a,ZhouW-2022a} is widespread, it is worth having a tool as reliable and easy to apply as MFDFA is in 1D analysis.  This is why the attempts were made to generalize the MFDFA and multifractal detrending moving average (MFDMA)~\cite{GuGF-2010a} formalisms beyond 1D, which involved studying the scaling properties of the residual fluctuation moments for a detrended 2D surface or image of interest~\cite{GuGF-2006a}.
 In fact, DMA was used even earlier in the context of generalizing the concept of Hurst exponents to many dimensions~\cite{CarboneA-2007}. There is some more recent numerical evidence that MFDFA is more efficient in terms of computational time than MFDMA, especially in more dimensions than one~\cite{XiC-2016}.

Although these methods were predisposed to capturing multi-scale characteristics in 2D in the average sense (like crossing the square of size $s \times s$ along the diagonal, thus using only one spatial variable and effectively projecting the problem on one dimension), its inherent symmetry regarding direction prevented it from detecting potential anisotropy of the fractal properties.

In the present approach, the above constraints are lifted by allowing for a rectangular grid of segments of size $s_{\rm x} \times s_{\rm y}$, where $s_{\rm x} \neq s_{\rm y}$ is allowed. By fixing $s_{\rm y}$ and varying $s_{\rm x}$, we break the isotropy and associate the multifractal analysis with a direction. The most extreme way to do this is to put $s_{\rm y}=1$ and to consider 1D strips of the surface cut along the X-axis direction as if they were 1D time series. Then such strips may be considered either separately or as the consecutive parts of a long time series. In the former, one may study the fractal properties of the fluctuations along each strip and observe how they depend on the Y-coordinate~\cite{AlvarezRamirezJ-2006a}. In the latter, one may study the average fractal properties along the X-axis direction without any dependence on the Y-coordinate. Of course, in a more general situation, one may put $s_{\rm y} > 1$ and investigate the fractal properties of wider strips. Finally, one may carry out another study, in which $s_{\rm y}$ varies over some range of scales. This can allow one for an assessment of whether fractal properties differ between different Y coordinates without the need to perform a 1D analysis for each strip separately. One can exploit here the result that, if neighbouring strips reveal different multifractal behaviour, their mixture (i.e., for $s_{\rm y}>1$) will tend to monofractality or bifractality~\cite{DrozdzS-2015a,KwapienJ-2023a}.

After defining a characteristic direction of analysis, the next logical step is to enable rotation of this direction and study the fractal properties of a given surface along different axes. This step enables studying surfaces that were subjected to directional processes and developed anisotropy. Our formalism allows adjustment of the rotation angle $\phi$ between 0 and 180 degrees, encompassing all possible directions. Thus, both the approach of ref.~\cite{GuGF-2006a} and the directional 1D MFDFA~\cite{AlvarezRamirezJ-2006a} may be viewed as special cases of the methodology presented here.

\section{Directional MFDFA in 2 dimensions}
\label{sect::methodology}

Let one consider a function $a: \mathbb{R}\times\mathbb{R} \rightarrow \mathbb{R}$ representing a scalar field over a 2D Euclidean space, which can be, for example, a 2D surface, or a digital image. For the sake of clarity, it will be referred to as a surface henceforth. Let it be sampled in such a way that it forms a rectangular grid of data points $a(m,n)$, where the grid is oriented along the X and Y axes and where $m=0,...,T_{\rm x}-1$ and $n=0,...,T_{\rm y}-1$ with $T_{\rm x},T_{\rm y} \gg 1$.  For a given choice of scale $s_{\rm x}$ in X-direction and $s_{\rm y}$ in Y-direction, one covers the data grid with $M_{s_{\rm x}} M_{s_{\rm y}}$ non-overlapping rectangular segments of size $s_{\rm x} \times s_{\rm y}$ labelled by $\bm{\nu}=(\nu_{\rm x},\nu_{\rm y}$), where $\nu_{\rm x}=0,...,M_{s_{\rm x}}-1$ and $\nu_{\rm y}=0,...,M_{s_{\rm y}}-1$. Here, $T_{\rm x}=s_{\rm x} M_{s_{\rm x}}$ and $T_{\rm y}=s_{\rm y} M_{s_{\rm y}}$ is assumed, but this requirement may be shifted according to the standard MFDFA procedure~\cite{KantelhardtJ-2002a}. To start with, let one assume that there is no rotation ($\phi=0$). In parallel to the standard 1D MFDFA, the subject of the analysis is not the surface itself, but rather its integrated surface profile $A_{\bm{\nu}}(k,l)$ calculated segmentwise:
\begin{equation}
A_{\bm{\nu}}(k,l) = \sum_{u=0}^k \sum_{w=0}^l a(\nu_{\rm x}s_{\rm x} + u,\nu_{\rm y}s_{\rm y} + w)
\label{eq::surface.profile}
\end{equation}
with $k=0,...,s_{\rm x}-1$ and $l=0,...,s_{\rm y}-1$. The next step is detrending of the segmented profile $A_{\bm{\nu}}(k,l)$, which is performed by subtracting a two-dimensional least-square-fitted polynomial surface $P_{\bm{\nu}}^{(r)}$ of degree $r$:
\begin{equation}
A^{\rm detr}_{\bm{\nu}}(k,l) = A_{\bm{\nu}}(k,l) - P_{\bm{\nu}}^{(r)}(k,l).
\end{equation}
For each segment $\bm{\nu}$, one calculates a 2D detrended variance:
\begin{equation}
f_{\rm A}^2 (\bm{s},\bm{\nu}) = {1 \over s_{\rm x} s_{\rm y}} \sum_{k=1}^{s_{\rm x}} \sum_{l=1}^{s_{\rm y}} \left[ A^{\rm detr}_{\bm{\nu}}(k,l) - \Tilde{A}^{\rm detr}_{\bm{\nu}} \right]^2,
\label{eq::local.variance}
\end{equation}
where $\bm{s} = (s_{\rm x},s_{\rm y})$ and $\Tilde{A}^{\rm detr}_{\bm{\nu}}$ is the mean value of $A^{\rm detr}_{\bm{\nu}}(k,l)$ in this segment. A $q$-dependent 2D fluctuation function is described by the following formula:
\begin{equation}
F_{\rm A}^q(\bm{s}) = \bigg\{ {1 \over M_{s_{\rm x}} M_{s_{\rm y}}} \sum_{\nu_{\rm x}=0}^{M_{s_{\rm x}}-1} \sum_{\nu_{\rm y}=0}^{M_{s_{\rm y}}-1} \left[ f_{\rm A}^2(\bm{s},\bm{\nu})\right]^{q/2} \bigg\}^{1/q}
\label{eq::fluctuation.function}
\end{equation}
for $q \neq 0$ and
\begin{equation}
F_{\rm A}^q(\bm{s}) = \exp \bigg\{ \frac{1}{2 M_{s_{\rm x}} M_{s_{\rm y}}} \sum_{\nu_{\rm x}=0}^{M_{s_{\rm x}}-1} \sum_{\nu_{\rm y}=0}^{M_{s_{\rm y}}-1} \ln f_{\rm A}^2(\bm{s},\bm{\nu}) \bigg\}
\end{equation}
for $q=0$. The subscript A in $F_{\rm A}^q$ indicates that one deals with the fluctuation function for 2D surface here in contrast to its counterpart for the standard 1D MFDFA. In the case of fractal objects, one observes a power-law dependence of $F_{\rm A}^q(\bm{s})$:
\begin{equation}
F_{\rm A}^q (\bm{s}) \sim (s_{\rm x}s_{\rm y})^{{1 \over 2} h_{\rm A}(q)}
\label{eq::scaling}
\end{equation}
with the exponent $h_{\rm A}(q)$ being a 2D version of the generalized Hurst exponent. In practical situations, one may prefer to fix $s_{\rm y}$ and consider only $s_{\rm x}$ as a variable; then it is possible to look for a relation of the form $F_{\rm A}^q (\bm{s}) \sim s_{\rm x}^{h_{\rm x}(q;s_{\rm y})}$.

\begin{figure*}
\begin{center}
\includegraphics[scale=0.37]{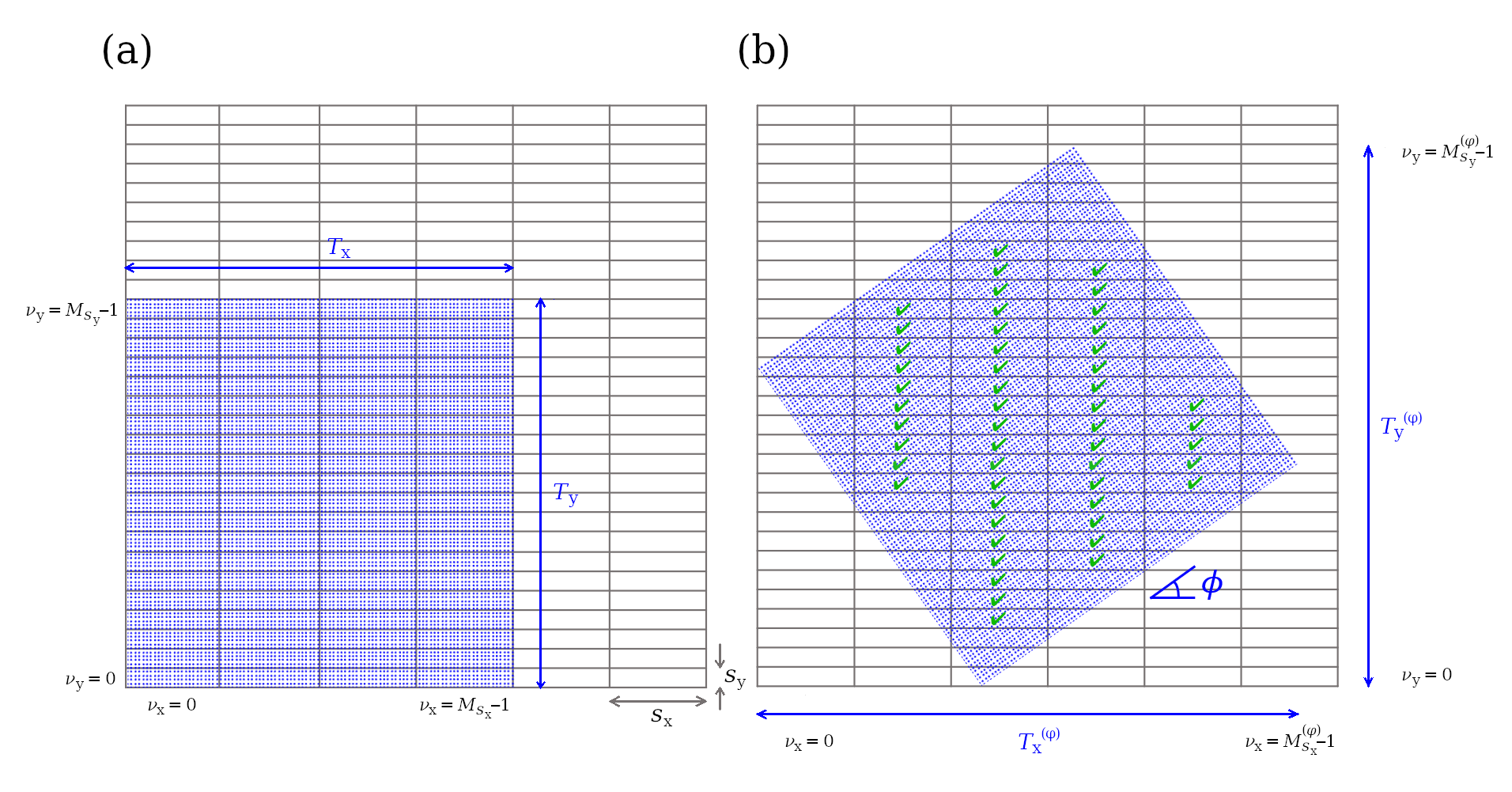}
\end{center}
\caption{The array of data points (blue dots) with the imposed regular grid of segments of size $s_{\rm x} \times s_{\rm y}$ (black lines). (a) The data array in its original position ($\phi = 0^{\circ}$) and (b) the same data rotated counterclockwise by a sample angle $\phi$. The segments that are considered in the $\phi \neq 0^{\circ}$ case have to be entirely contained within the data array (green check marks).}
\label{fig::method.illustration}
\end{figure*}

Now, instead of considering a static data grid that is oriented along the X-Y axes, we allow for its counterclockwise rotation by an angle $0^{\circ} < \phi \le 180^{\circ}$ in the X-Y plane. The same effect can be obtained if one rotates the grid of the cover segments by an angle $-\phi$, but for the sake of computational simplicity, a data rotation is preferred (see Fig.~\ref{fig::method.illustration}). The coordinates of each point of the data grid are transformed as $a(m,n) \longrightarrow a^{\phi}(g_{\phi}(m),h_{\phi}(n))$ according to the following rules:
\begin{alignat}{2}
\nonumber
g_{\phi}(m) & = m \cos\phi - n \sin\phi + (T_{\rm y}-1) \sin\phi + \\
& \qquad\qquad\qquad\qquad - \Theta (\phi-\frac{\pi}{2})(T_{\rm x}-1) \cos\phi,\\
\nonumber
h_{\phi}(n) & = m \sin\phi + n \cos\phi - \Theta(\phi-\frac{\pi}{2})(T_{\rm y}-1) \cos\phi,
\end{alignat}
where $T_{\rm x}$ and $T_{\rm y}$ is the number of grid points along the X-axis and Y-axis, respectively. $\Theta(x)$ is the Heaviside step function that is equal to 1 if $x \ge 0$ or 0 if $x < 0$ and related terms have been added to avoid negative coordinates of the grid points after the transformation. These new coordinates have the following bounds:
\begin{equation}
g_{\phi}(m) \in [0,T_{\rm x}^{(\phi)}], \qquad h_{\phi}(n) \in [0,T_{\rm y}^{(\phi)}],
\end{equation}
where
\begin{eqnarray}
\nonumber
T_{\rm x}^{(\phi)} = (T_{\rm y}-1)\sin\phi + {\rm sgn}(\frac{\pi}{2} - \phi) (T_{\rm x}-1)\cos\phi,\\
T_{\rm y}^{(\phi)} = (T_{\rm x}-1)\sin\phi + {\rm sgn}(\frac{\pi}{2} - \phi) (T_{\rm y}-1)\cos\phi.
\end{eqnarray}
(Note that $T_{\rm x}^{(\phi)},T_{\rm y}^{(\phi)}$ are real numbers depending on $\phi$.)

Because the segment grid is fixed in its original orientation and only the surface is subject to rotations, the size of individual segments is the same as before, $s_{\rm x} \times s_{\rm y}$, but their number must be larger to cover the entire rotated surface: $M_{s_{\rm x}}^{(\phi)}=\lfloor T_{\rm x}^{(\phi)} / s_{\rm x} \rfloor$ and $M_{s_{\rm y}}^{(\phi)}=\lfloor T_{\rm y}^{(\phi)} / s_{\rm y} \rfloor$. Thus, $\nu_{\rm x} = 0,...,M_{s_{\rm x}}^{(\phi)}-1$ and $\nu_{\rm y} = 0,...,M_{s_{\rm y}}^{(\phi)}-1$. Because the grid and the data rows and columns are not aligned in the present case, each segment $\bm{\nu}$ may contain a different number $N_{\bm{\nu}}$ of data points, including the possibility that there exist empty segments outside the surface (see Fig.~\ref{fig::method.illustration}(b)). The MFDFA algorithm does not allow for the consideration of segments with the same size but with significantly different number of data points inside; therefore, only those segments are taken into consideration, which are located entirely within the bounds of the studied surface. This ensures that only minor fluctuations in the number of data points in each segment can be expected. It follows that the number of segments that can be considered depends on $\phi$. (Note that with this approach, for large segment sizes, many data points may fall outside the segments taken into account. In such a situation, the maximum values of $s_{\rm x}$ and $s_{\rm y}$ should be limited because the smaller they are, the fewer points remain unconsidered. Another possible approach to avoid this problem is to apply a stochastic sampling, i.e., to consider segments whose position was randomly selected rather than having a grid of segments with fixed positions. If the number of such random segments is large enough, almost the whole data array can be penetrated in this way, and only a few data points are left without consideration even if the segments are relatively large.)

In the present case, a residual detrended surface is given by
\begin{equation}
A^{\phi,\rm detr}_{\bm{\nu}}(i) = A^{\phi}_{\bm{\nu}}(i) - P_{\bm{\nu}}^{(r)}(i),
\end{equation}
where $i=1,...,N_{\bm{\nu}}$ (the order of labelling points $i$ on the surface may be arbitrary). In analogy to Eq.~(\ref{eq::local.variance}), the 2D detrended directional variance is given by
\begin{equation}
f_{\rm A}^2 (\bm{s},\bm{\nu};\phi) = {1 \over N_{\bm{\nu}}} \sum_{i=1}^{N_{\bm{\nu}}} \left[ A^{\phi,\rm detr}_{\bm{\nu}}(i) - \Tilde{A}^{\phi,{\rm detr}}_{\bm{\nu}} \right]^2,
\label{eq::local.variance.rotated}
\end{equation}
and, in analogy to Eq.~(\ref{eq::fluctuation.function}), the $q$-dependent 2D directional fluctuation function for surface $A^{\phi}$ is given by
\begin{equation}
F_{\rm A}^q(\bm{s};\phi) = \bigg\{ {1 \over M_{*}^{(\phi)}} \sum_{\bm{\nu}_{*}} \left[ f_{\rm A}^2(\bm{s},\bm{\nu};\phi)\right]^{q/2} \bigg\}^{1/q}
\label{eq::fluctuation.function.rotated}
\end{equation}
for $q \neq 0$ and
\begin{equation}
F_{\rm A}^q(\bm{s};\phi) = \exp \bigg\{ \frac{1}{2 M_{*}^{(\phi)}} \sum_{\bm{\nu}_{*}} \ln f_{\rm A}^2(\bm{s},\bm{\nu};\phi) \bigg\}
\label{eq::fluctuation.function.rotated}
\end{equation}
for $q=0$. The summation in the above formula takes place only over such segments $\bm{\nu}_{*}$ that do not extend beyond the surface boundary; their total number is denoted by $M^{(\phi)}_{*}$. Obviously, Eqs.~(\ref{eq::local.variance.rotated}) and (\ref{eq::fluctuation.function.rotated}) reduce to Eqs.~(\ref{eq::local.variance}) and (\ref{eq::fluctuation.function}), respectively, for $\phi=0$.

It may happen that one is interested in the strip-wise properties of the studied surface in direction $\phi$ rather than its global properties. If this is the case, one may introduce the strip-wise directional fluctuation functions defined by
\begin{equation}
_{\rm _X}F_{\rm A}^q(\bm{s},\nu_{\rm y};\phi) = \bigg\{ {1 \over M^{(\phi)}_{\rm x *}} \sum_{\nu_{\rm x *}} \left[ f_{\rm A}^2(\bm{s},\bm{\nu};\phi)\right]^{q/2} \bigg\}^{1/q}
\label{eq::fluctuation.function.rotated.x}
\end{equation}
for $q \neq 0$ and
\begin{equation}
_{\rm _X}F_{\rm A}^q(\bm{s},\nu_{\rm y};\phi) = {1 \over M^{(\phi)}_{\rm x *}} \sum_{\nu_{\rm x *}} \ln f_{\rm A}^2(\bm{s},\bm{\nu};\phi)
\label{eq::fluctuation.function.rotated.x}
\end{equation}
for $q=0$, which can be calculated for each segment row $\nu_{\rm y}$ separately. If they are then averaged over all the $\nu_{\rm y}$ strips, one arrives at the approach proposed in Ref.~\cite{AlvarezRamirezJ-2006a}. (Note that the same can be done in the perpendicular direction if one replaces $\phi$ with $\phi+90^{\circ}$, so there is no need to define a Y-axis counterpart of Eq.~(\ref{eq::fluctuation.function.rotated.x}).)

The surface structure may be considered as fractal if
\begin{equation}
F_{\rm A}^q (\bm{s};\phi) \sim (s_{\rm x}s_{\rm y})^{{1 \over 2} h_{\rm A}(q;\phi)}
\label{eq::scaling.rotated}
\end{equation}
with the exponent $h_{\rm A}(q)$ being a 2D version of the generalized Hurst exponent. Again, Eq.~(\ref{eq::scaling.rotated}) may be replaced by
\begin{equation}
F^q_{\rm A}(\bm{s};\phi) \sim s_{\rm x}^{h_{\rm x}(q;s_{\rm y},\phi)}
\label{eq::scaling.rotated.fixed.y}
\end{equation}
if $s_{\rm y}$ has been fixed.

A more concise way of quantifying the fractal properties of an object is through a singularity spectrum, which shall be denoted here by $\mathcal{F}$ in order to avoid confusion with variance $f_{\rm A}^2$:
\begin{eqnarray}
\nonumber
\alpha^{(\phi)}_{\rm A} = h_{\rm A}(q;\phi) + q h_{\rm A}'(q;\phi),\\
\mathcal{F}(\alpha^{(\phi)}_{\rm A}) = q \left[ \alpha^{(\phi)}_{\rm A} - h_{\rm A}(q;\phi) \right] + D,
\label{eq::singularity.spectrum}
\end{eqnarray}
where $\alpha^{(\phi)}_{\rm A}$ denotes the H\"older exponent and $'$ denotes the first derivative with respect to $q$. The parameter $D$ is the topological dimension of the data set support, therefore $D=2$ for $h_{\rm A}(q;\phi)$ and $D=1$ for $h_{\rm x}(q;s_{\rm y},\phi)$.

\section{Model and empirical data}
\label{sect::model.empirical.data}

\subsection{Exemplary model processes with known properties}

Performance of the method is tested using a few model fractal processes on a 2D surface represented by a square array.

\begin{figure}
\begin{center}
\includegraphics[scale=0.55]{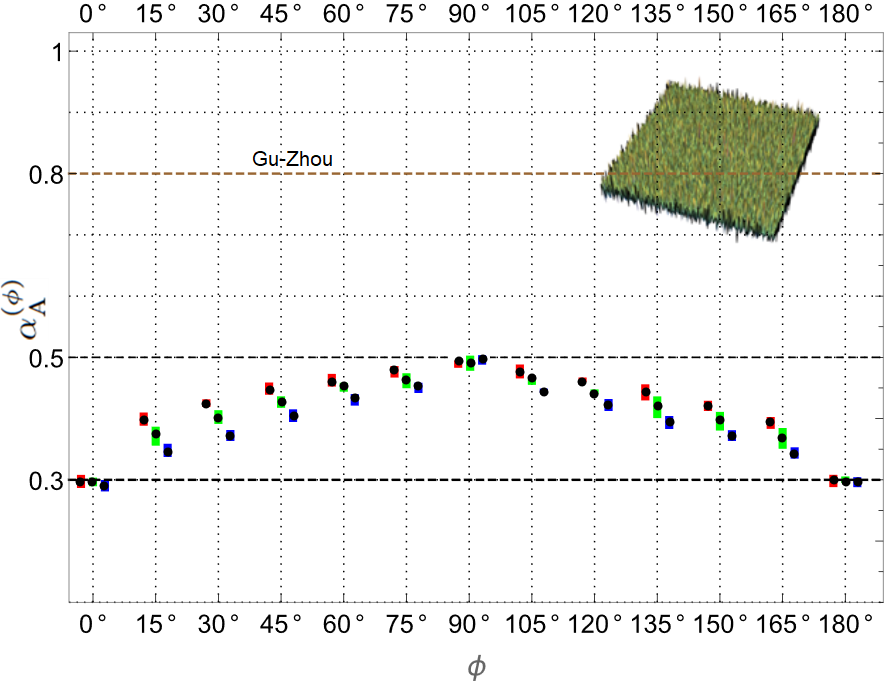}
\end{center}
\caption{Directional MFDFA in 2D applied to a surface defined by a $2,000 \times 2,000$ array (inset) whose rows contain fractional Gaussian noise with $H=0.3$; each row represents an independent realization of the process. $\mathcal{F}(\alpha^{(\phi)}_{\rm A})$ for a given direction $0^{\circ} \le \phi \le 180^{\circ}$ for three sample values of the (fixed) segment size in Y-axis, $s_{\rm y}$, are shown as vertical projections on the $\alpha$-axis. The maximum $\mathcal{F}(\alpha^{(\phi)}_{\rm A})$ is denoted by a black dot and the variability range of $\alpha^{(\phi)}_{\rm A}$ for $-10 \leqslant q \leqslant 10$ by vertical color strips: red ($s_{\rm y}=30$), green ($s_{\rm y}=80$), and blue ($s_{\rm y}=200$), slightly shifted horizontally to improve readability. $\phi=0^{\circ}$ corresponds to the direction of the array rows, while $\phi=90^{\circ}$ to the array columns. A horizontal line at $\alpha=0.8$ illustrates the result of direction-insensitive approach~\cite{GuGF-2006a}.}
\label{fig::Gaussian.noise}
\end{figure}

{\bf Monofractal case.} First, a fractional Gaussian noise with the sample Hurst exponent $H=0.3$ is generated independently in each row of the array~\cite{Mandelbrot-1968}. The consecutive elements in each column thus contain uncorrelated noise with $H=0.5$. This leads to a $\phi$-dependence of the fractal properties as shown in Fig.~\ref{fig::Gaussian.noise}. The length of the color vertical strips indicating the width of the $\mathcal{F}(\alpha^{(\phi)}_{\rm A})$ spectrum is so small that it suggests monofractal scaling of the data (i.e., no $q$-dependence of $h_{\rm A}(q;\phi)$) independently of $\phi$. What does change, however, is the value of $\alpha^{(\phi)}_{\rm A}$ as a function of $\phi$: for $\phi=0^{\circ}$ (the X-axis direction) it equals $\approx 0.3$, while it increases to $\approx 0.5$ for $\phi=90^{\circ}$ (the Y-axis direction). In contrast, $\alpha_{\rm A} \approx 0.8$ in the approach proposed in~\cite{GuGF-2006a}. Note that $0.8=0.3+0.5$, which shows that it is possible to extract extra genuine information by using the new methodology.

\begin{figure}
\begin{center}
\includegraphics[scale=0.75]{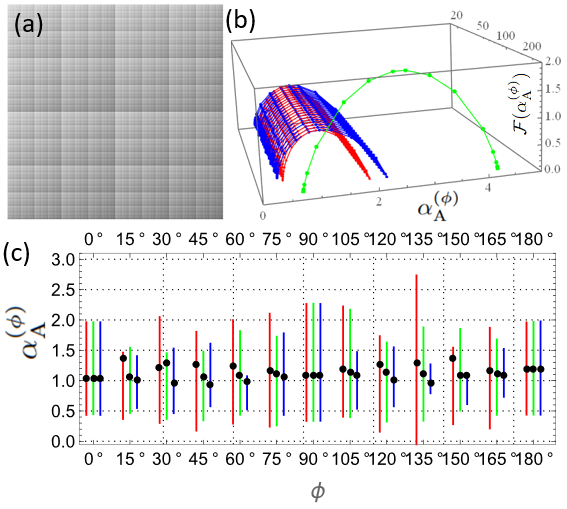}

\vspace{0.1cm}
\includegraphics[scale=0.75]{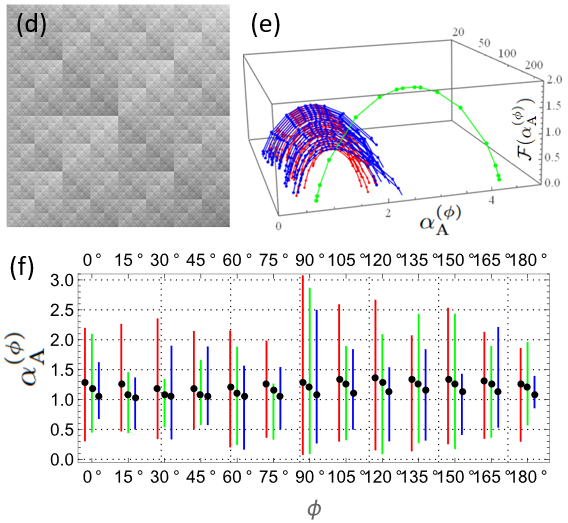}
\end{center}
\caption{10-level-deep multiplicative cascades of size $2\times 2$ with the weights: (a)-(c) 0.05, 0.15, 0.2, 0.6 and (d)-(f) 0.2, 0.05, 0.15, 0.6. (a),(d): $1024 \times 1024$ array that contains the corresponding cascades. (b),(e): singularity spectra $\mathcal{F}(\alpha^{(\phi)}_{\rm A})$ for $\phi=0^{\circ}$ (red) and $\phi=90^{\circ}$ (blue) together with the one for the approach (green) of ref.~\cite{GuGF-2006a}. (c),(f): vertical projection of $\mathcal{F}(\alpha^{(\phi)}_{\rm A})$ on $\alpha$-axis for a given direction $0^{\circ} \le \phi \le 180^{\circ}$ and sample segment size $s_{\rm y}$: 30 (red), 80 (green), and 200 (blue). The maximum $\mathcal{F}(\alpha^{(\phi)}_{\rm A})$ is denoted by a black dot, while the range of $q$ is $-5 \leqslant q \leqslant 5$.}
\label{fig::multiplicative.cascades}
\end{figure}

{\bf Multifractal case.} A generically multifractal process in 2 dimensions can be represented by a multiplicative cascade of depth $k \gg 1$ defined in a $2 \times 2$ square lattice with weights $\omega_{ij}$, where $i,j=0,1$ -- see Fig.~\ref{fig::multiplicative.cascades}. Two such cascades are considered, differing from each other only in their weight order (compare the visualizations of Fig.~\ref{fig::multiplicative.cascades} (a) and (d)). Although both cascades produce roughly the same singularity spectrum when the method of ref.~\cite{GuGF-2006a} is applied (a green curve in Fig.~\ref{fig::multiplicative.cascades} (b),(e)), there is a clear difference in $\mathcal{F}(\alpha^{(\phi)}_{\rm A})$ between the cascades with respect to $s_{\rm y}$ if $\phi=0^{\circ},90^{\circ},180^{\circ}$ (i.e., the directions coincide with the rows and columns of the data array). In this case, $\mathcal{F}(\alpha^{(\phi)}_{\rm A})$ is independent of $s_{\rm y}$ for the cascade depicted in Fig.~\ref{fig::multiplicative.cascades}(a), while it varies with $s_{\rm y}$ for the cascade depicted in Fig.~\ref{fig::multiplicative.cascades}(d) -- compare Fig.~\ref{fig::multiplicative.cascades}(b)(c) with Fig.~\ref{fig::multiplicative.cascades}(e)(f). The $\phi$-invariance observed for the cascade (a) vanishes if the angle differs from these three values and the width of $\mathcal{F}(\alpha^{(\phi)}_{\rm A})$ depends on $\phi$ now. There are also significant differences in the width of the spectra between both cascades for the same $\phi$ -- compare Fig.~\ref{fig::multiplicative.cascades} (c) and (f). (Note that fixing of the $s_{\rm y}$ values implies that Eq.~(\ref{eq::scaling.rotated.fixed.y}) has to be applied and $D=1$ in Eq.~(\ref{eq::singularity.spectrum}), while the approach of ref.~\cite{GuGF-2006a} is associated with using Eq.~\ref{eq::scaling.rotated} and $D=2$. This results in the clear difference between the maximum values of $\mathcal{F}(\alpha^{(\phi)}_{\rm A})$ if the green curve is compared with the blue and red ones in Fig.~\ref{fig::multiplicative.cascades}(b)(e).)

The widths of $\mathcal{F}(\alpha^{(\phi)}_{\rm A})$ in Fig.~\ref{fig::multiplicative.cascades} have tendency to decrease with increasing $s_{\rm y}$ irrespective of the grid orientation $\phi$. This is because, for $s_{\rm y}>1$, the method mixes different rows of the data array, which correspond to different organization of the cascade values. In order to give an even more explicit example, a set of 1,000 mutually correlated 1D binomial cascades has been generated. The strength of cross-correlations between processes has been controlled by a ratio $\sigma$ of multipliers that are changed at consecutive levels of the cascades. Consequently, each realization of the process forms a row in an array, making the data multifractal in the X-direction and cross-correlated in the Y-direction. It is a known effect that the calculation of the fractal properties of the signals that mix different multifractal processes depends on the number of such processes: mixing alters the inner organization of each process, and this leads to suppression of the multifractality, which is based on long-range correlations that are destroyed the more, the larger number of signals have been mixed~\cite{DrozdzS-2015a}.

\begin{figure}[H]
\begin{center}
\includegraphics[scale=0.92]{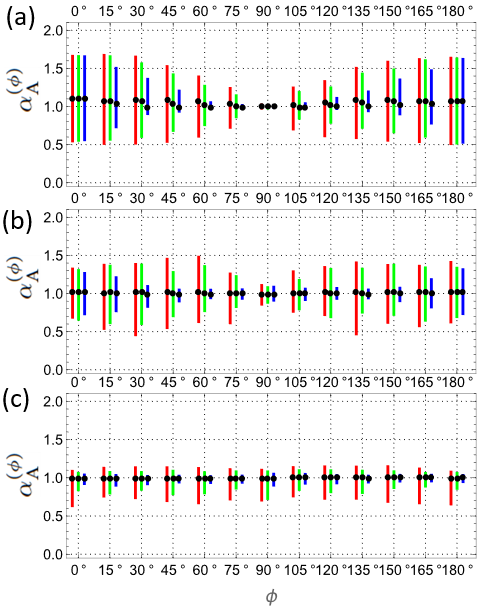}
\end{center}
\caption{Binomial multiplicative cascades with correlated multipliers (array rows, $\phi=0$) forming an array of size $1000 \times 1024$. The larger the level $\sigma$ is, the less correlated are the adjacent array rows: (a) $\sigma=0.1$, (b) $\sigma=0.5$, (c) $\sigma=0.9$. The plots show variability range of $\alpha^{(\phi)}_{\rm A}$ in the singularity spectra $\mathcal{F}(\alpha^{(\phi)}_{\rm A})$ (color vertical lines) obtained with the directional MFDFA in 2D for different orientations $\phi$ and three sample values of $s_{\rm y}$: 30 (red), 80 (green), and 200 (blue). The lines for the same $\phi$ and different $s_{\rm y}$ have been shifted horizontally relative to each other to improve readability. The maximum $\mathcal{F}(\alpha^{(\phi)}_{\rm A})$ is denoted by a black dot.}
\label{fig::correlated.cascades}
\end{figure}

Fig.~\ref{fig::correlated.cascades} shows the widths of $\mathcal{F}(\alpha^{(\phi)}_{\rm A})$ for three values of $\sigma$ decreasing from top (a) to bottom (c) and three values of $s_{\rm y}$, which is equal here to the number of cascades that are mixed within each strip of segments. By extending $s_{\rm y}$, one averages the variance functions (Eq.~(\ref{eq::local.variance.rotated})) over data points from an increasing number of different processes. Additionally, as the cross-correlation between the rows decreases with decreasing $\sigma$, the multifractality of the overall structure weakens. It happens even in the direction of the cascades (X-direction), therefore one observes here a decreasing dependence of the fractal properties on $\phi$. It should be stressed here that such effects could not be identified if the methodology of ref.~\cite{GuGF-2006a} was applied.

\subsection{Sample empirical data sets}

Since the methodology works as intended, the focus moves to sample sets of empirical data in order to illustrate how the new methodology works in practice. What can be looked for in this respect are possible signatures of a fractal structure in each set and their dependence on direction. The methodology described in Sect.~\ref{sect::methodology} is applied step by step for $0^{\circ} \le \phi \le 180^{\circ}$ and $-10 \le q \le 10$. A relatively broad range of scales is considered $20 \le s_{\rm x},s_{\rm y} \le 200$, but only three sample scales $s_{\rm y}$ are discussed in each case to show how the obtained results depend on $s_{\rm y}$.

\begin{figure}[H]
\begin{center}
\includegraphics[scale=0.55]{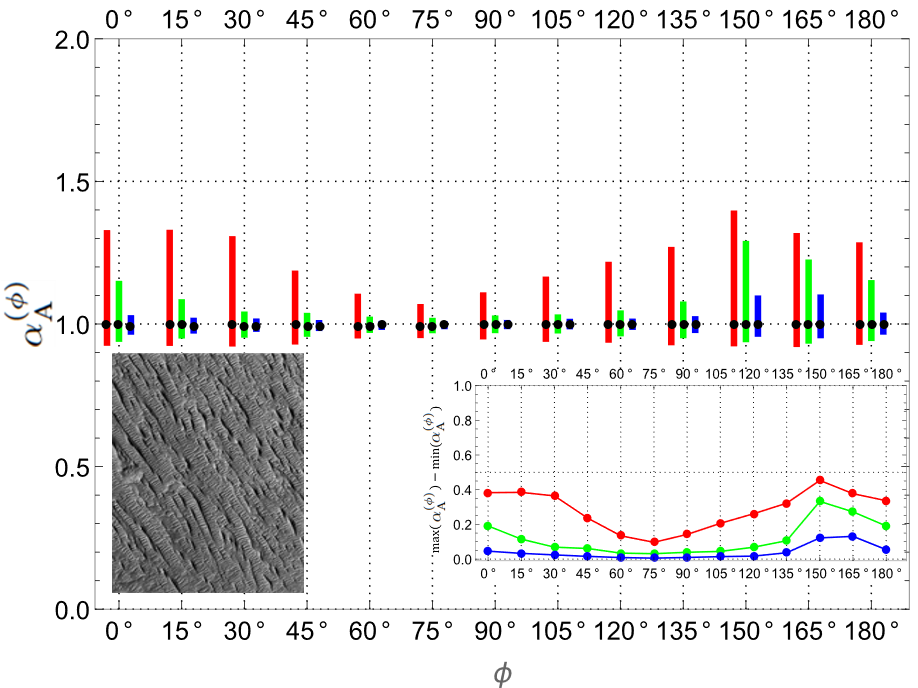}
\end{center}
\caption{(Bottom left) $1076 \times 672$ pixel gray-scale image of a Mars surface fragment (source: NASA JPL~\cite{Mars}). (Bottom right) The width of the singularity spectrum $\mathcal{F}(\alpha^{(\phi)}_{\rm A})$ as a function of the orientation $\phi$ for $-10 \le q \le 10$ and for different values of $s_{\rm y}$: 25 (red), 60 (green), and 180 (blue). (Main) Variability range of $\alpha^{(\phi)}_{\rm A}$ in $\mathcal{F}(\alpha^{(\phi)}_{\rm A})$ as a function of $\phi$ (colors as above). The maximum $\mathcal{F}(\alpha^{(\phi)}_{\rm A})$ is denoted by a black dot.}
\label{fig::mars}
\end{figure}

{\bf Mars surface.} The first data set is a digital image of a region of the Mars surface~\cite{Mars}, which shows some directional anisotropy visible to the naked eye -- see Fig.~\ref{fig::mars} (top left). The singularity spectra $\mathcal{F}(\alpha^{(\phi)}_{\rm A})$ are mainly right-side asymmetric (i.e., the black dots denoting the spectrum maxima are placed below the middle of the lines representing the projections of $\mathcal{F}(\alpha^{(\phi)}_{\rm A})$ on the $\alpha^{(\phi)}$ axis). Such an asymmetry indicates that these are the large data values, which show rich multifractality, while the small values are more noisy and tend towards monofractal behaviour~\cite{DrozdzS-2015a}. The results confirm the presence of anisotropy: for $\phi \approx 75^{\circ}$ (which roughly overlaps with the direction along the furrows in the digital image in Fig.~\ref{fig::mars}), the surface is essentially monofractal, while it acquires the multifractal property if $\phi$ moves away from this angle with the significant multiscaling observed for $0^{\circ} \leqslant \phi \leqslant 30^{\circ}$ and $120^{\circ} \leqslant \phi \leqslant 180^{\circ}$ peaking at $\phi=150^{\circ}$. The maximum of $\mathcal{F}(\alpha^{(\phi)}_{\rm A})$ is independent of $\phi$ and indicates a significant overall smoothness of the surface.

\begin{figure}[H]
\begin{center}
\includegraphics[scale=0.55]{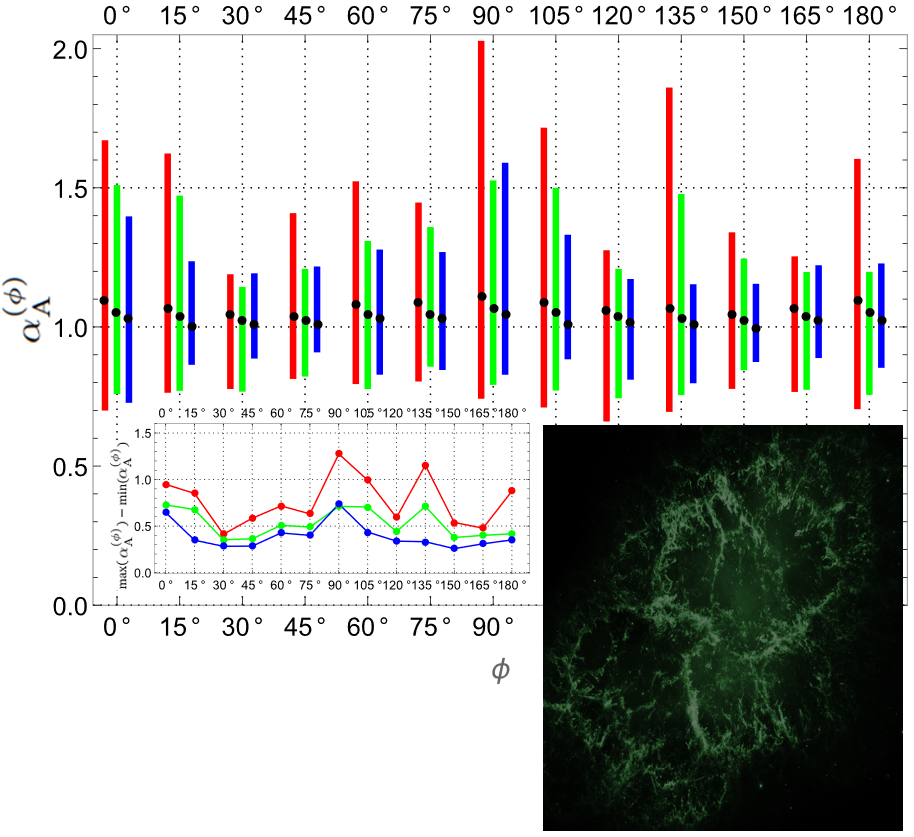}
\end{center}
\caption{Digital image of the Crab Nebula (M1) made by Hubble Space Telescope in visible light ($1920 \times 1080$ pixels, true colors reduced to gray scale, source: NRAO~\cite{Crab}). The same quantities as in Fig.~\ref{fig::mars}.}
\label{fig::crab.nebula}
\end{figure}

{\bf Crab Nebula.} A digital image of the Crab Nebula obtained in the visible light spectrum~\cite{Crab}, which exhibits certain marks of anisotropy, has also been subject to the directional MFDFA in 2D -- see the top image in Fig.~\ref{fig::crab.nebula}. Indeed, the singularity spectra $\mathcal{F}(\alpha^{(\phi)}_{\rm A})$, although always multifractal, present also a strong dependence on direction. The richest multiscaling is seen for $\phi=0^{\circ}$, $90^{\circ}$, and $135^{\circ}$, while the poorest one for $\phi=30^{\circ}$ and $150^{\circ}$. By looking carefully at the empirical image, the latter angles can be associated with the directions of lighter irregular streaks, which account for the observed variability of the spectra. One has to keep in mind, however, that the analyzed image of the Crab Nebula is actually a 2D projection of a 3D object on the camera image sensors, so much information about its actual geometry has already been lost. A complete study of such an object would require a 3D approach, which is easy to be designed by generalizing the present 2D method to 3D along the same line as that presented in Sect.~\ref{sect::methodology}, if only the respective data sets were available.

\begin{figure}[H]
\begin{center}
\includegraphics[scale=0.38]{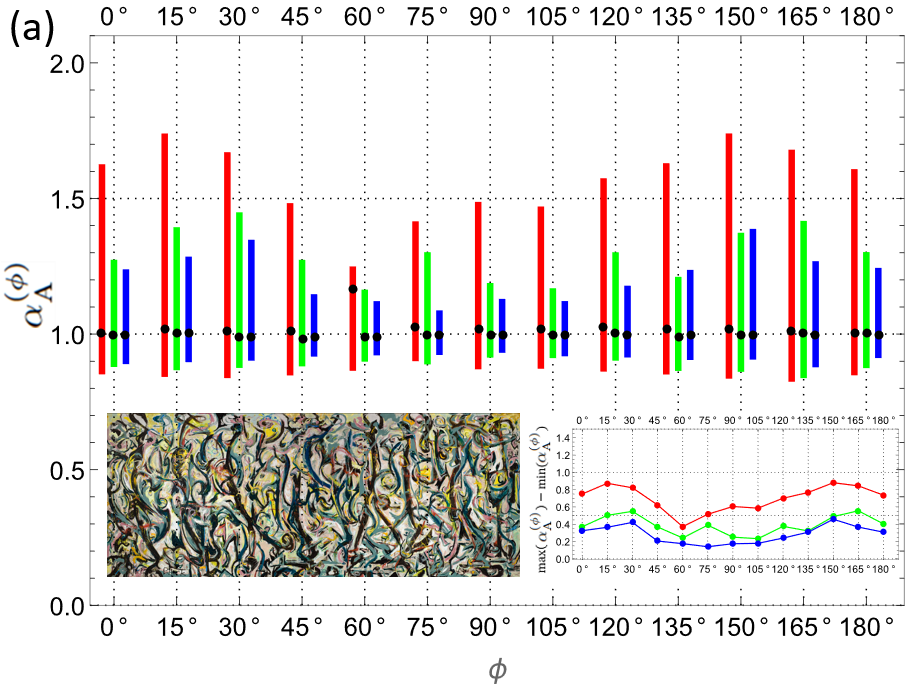}
\vspace{0cm}
\includegraphics[scale=0.38]{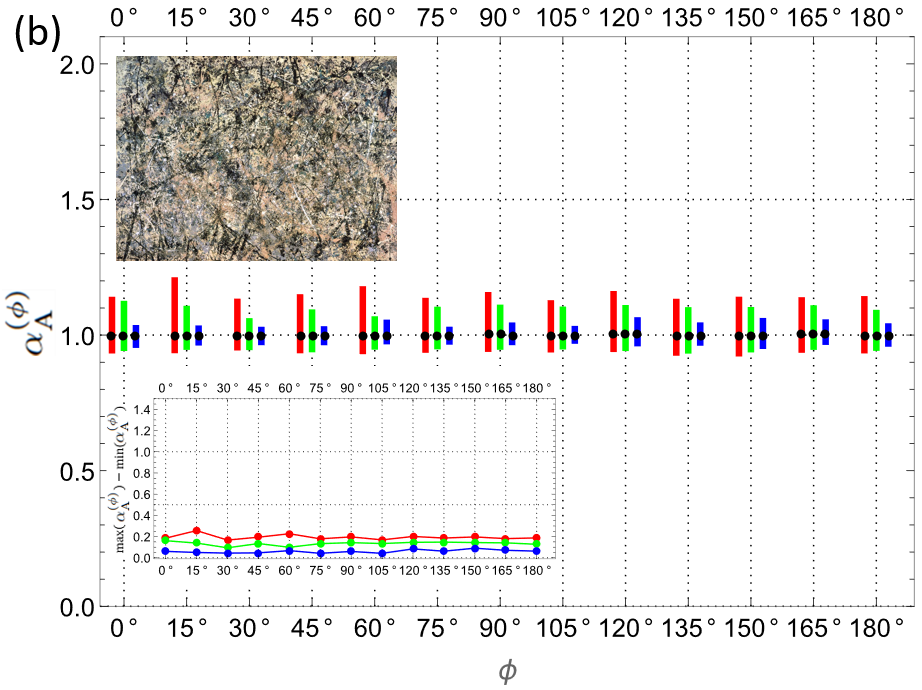}
\vspace{0cm}
\includegraphics[scale=0.38]{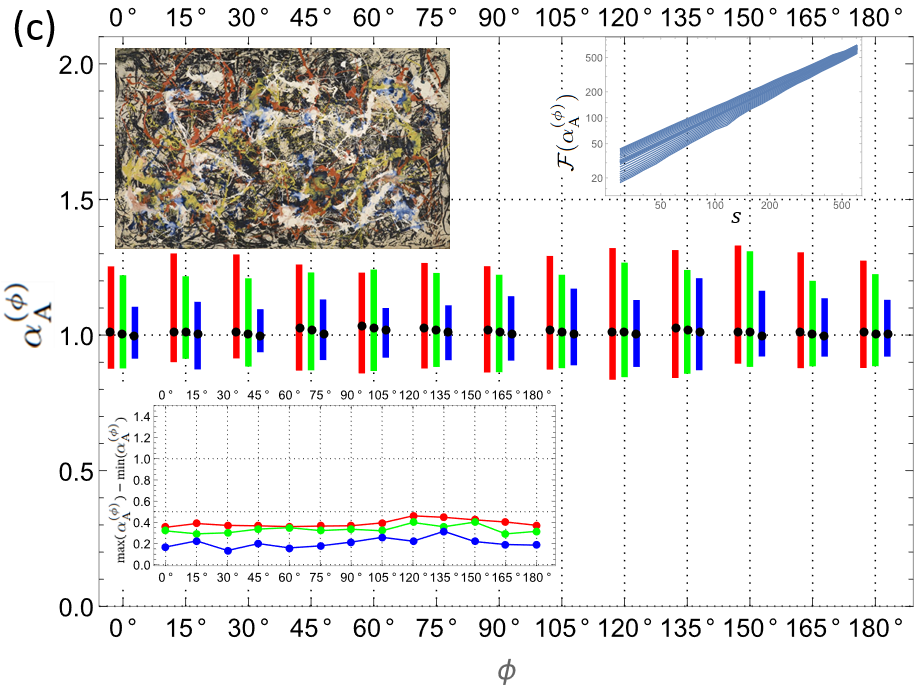}
\end{center}
\caption{Digital images of Jackson Pollock's paintings (source: Pollock Painting Gallery~\cite{Pollock}): (a) \textit{Mural} (1943) representing early works ($963 \times 2429$ pixels), (b) \textit{Lavender Mist} (1950) representing peak-career works ($1365 \times 2048$ pixels), and (c) \textit{Convergence} (1952) representing late, retrospective works ($1200 \times 2000$ pixels). Each plot shows the painting, the variability range of $\alpha^{(\phi)}_{\rm A}$ in $\mathcal{F}(\alpha^{(\phi)}_{\rm A})$ as a function of $\phi$ (main panel) and the width of $\mathcal{F}(\alpha^{(\phi)}_{\rm A})$ also as a function of $\phi$ (inset). The color codes and the meaning of the black dots are the same as in Figs.~\ref{fig::mars} and~\ref{fig::crab.nebula}. In addition, the inset of the bottom plot presents a typical fluctuation function $F_{\rm A}^q(\bm{s};\phi)$, here for $\phi=0$ and $s_{\rm y}=25$.}
\label{fig::pollock}
\end{figure}

{\bf Jackson Pollock's paintings.} The final empirical examples refer to Jackson Pollock's paintings~\cite{Pollock} representing different sub-periods of his action-painting period: the early one (\textit{Mural}, 1943 -- a work that still exhibits signs of the former abstraction period), the peak one (\textit{Convergence}, 1952), and the late one (\textit{Lavender Mist}, 1950). The works of Pollock from the action-painting period are known to reveal fractal patterns~\cite{TaylorRP-1999a,AlvarezRamirezJ-2008a,MureikaJR-2013a} whose multifractality became poorer in later periods. Therefore, it is instructive to compare these works also from the perspective of direction. Of the three examples, no doubt it is \textit{Mural} that shows the richest multifractality (Fig.~\ref{fig::pollock}(a)), which is also substantially $\phi$-dependent with the maximum multiscaling detected for $15^{\circ} \leqslant \phi \leqslant 30^{\circ}$ and $150^{\circ} \leqslant \phi \leqslant 165^{\circ}$ and the minimum for $60^{\circ} \leqslant \phi \leqslant 105^{\circ}$. \textit{Lavender Mist} represents the most acclaimed period of Pollock's career, when he used to apply his drip technique. It shows a substantially different character: a much more isotropic nature with less developed or even absent multifractality, characterized by only a minor increase in the width of $\mathcal{F}(\alpha^{(\phi)}_{\rm A})$ mainly for $\phi=15^{\circ}$ and $60^{\circ}$ (Fig.~\ref{fig::pollock}(b)).

\textit{Convergence}, which was created later than \textit{Lavender Mist} and after the end of the peak career period, is of particular interest, because it may be viewed as a retrospective of the earlier periods, unifying the dark, more figurative features typical for the early period and the multilayer color pouring typical for the peak period~\cite{KantorJ-2016a}. Consistently, results of the multifractal analysis may be characterized as intermediate in both the width of $\mathcal{F}(\alpha^{(\phi)}_{\rm A})$ and directionality (Fig.~\ref{fig::pollock}(c)): the width fluctuations are more prominent than in the case of \textit{Lavender Mist} yet significantly smaller than in the case of \textit{Mural}. The principal directions of richer multifractality are in the range $120^{\circ} \leqslant \phi \leqslant 150^{\circ}$. It is worth mentioning that the fluctuation functions $F^q_{\rm A}(\bm{s};\phi)$ for \textit{Convergence} show multifractal scaling with good quality - see the top-right inset in Fig.~\ref{fig::pollock}(c). The difference in the magnitude of anisotropy among the paintings from different periods can easily be explained by the visible broken symmetry of the earlier works and the largely symmetric geometry of the later ones. Likely, this variation stems from Pollock's evolving technique over time as he incorporated more colors and layers into his work.

\section{Conclusions}
\label{sect::conclusions}

A generalization of MFDFA formalism that allows one to selectively focus on a distinguished spatial direction in 2D, defined by an angle $\phi$, has been proposed. The resulting directional MFDFA in 2D includes thus the two former approaches reported in Refs.~\cite{AlvarezRamirezJ-2006a,GuGF-2006a} as its special cases. The method has been proven useful for identification of anisotropy in the fractal properties of 2D objects like, for instance, rough surfaces and digital images. The reliability of the methodology is demonstrated on model datasets, enabling its application to empirical data with unknown fractal properties. In particular, images of a Mars surface sample and the Crab Nebula reveal direction-dependent multifractality. This is also the case of works by Jackson Pollock made with the drip technique. The analysis using the method proposed here indicates that multifractal characteristics apply to Pollock's works, and their changes in terms of intensity and degree of anisotropy go parallel to his modifications of the painting technique. It has also been verified by direct calculation that in all the cases discussed, the destruction of correlations by a random redistribution of values on the surfaces under consideration leads to the disappearance of multifractality by reducing the scaling to monofractal. This indicates that, as in the 1D case~\cite{DrozdzS-2009,Zhou-2012,DrozdzS-2015a,KwapienJ-2023a}, the observed multifractality in 2D originates from correlations.

The three examples of the application of the proposed method for capturing asymmetry effects in the multifractal organization of 2D patterns presented here are only a small part of the potential applicability of the method. The usefulness of this method may even apply to quantum phenomena such as Anderson localization, metal-insulator transitions~\cite{Mirlin-2000} and to asymmetric quantum billiards~\cite{Lozej-2022}. Finally, the method presented here can be also applied using other methods of detrending like moving average~\cite{GuGF-2010a} and even its 2D generalization of the existing 1D methods like MF-DXA~\cite{Zhou-2008}, MF-X-DMA~\cite{Jiang-2011}, MF-CCA~\cite{Oswiecimka-2014}, MF-X-WT~\cite{Jiang-2017a}, MF-W-XL~\cite{Jiang-2017b} for examining multifractal cross-correlations can be considered.

\section*{Data Availability Statement}

The computer-generated datasets analysed during the current study are available from the corresponding author on reasonable request, while the empirical datasets are available from the cited repositories: NASA JPL (Mars surface images)~\cite{Mars}, NRAO (Crab Nebula images)~\cite{Crab}, and Pollock Painting Gallery (digital images of Pollock's paintings)~\cite{Pollock}.

\end{document}